\documentclass[doublecol]{epl2}
\usepackage{graphicx}
\usepackage{dcolumn}
\usepackage{bm}

\title{Spatial clustering of interacting bugs: \\L\'evy flights versus Gaussian jumps}

\author{E.~Heinsalu\inst{1,2} \and E.~Hern\'andez-Garc\'ia\inst{1} \and C.~L\'opez\inst{1}}
\shortauthor{E.~Heinsalu \etal}

\institute{
  \inst{1} IFISC, Instituto de F\'isica Interdisciplinar y Sistemas Complejos (CSIC-UIB),
  E-07122 Palma de Mallorca, Spain\\
  \inst{2} National Institute of Chemical Physics and Biophysics,
  R\"avala 10, Tallinn 15042, Estonia
}

\abstract{
A biological competition model where the individuals of the same species perform
a two-dimensional Markovian continuous-time random walk and undergo reproduction and
death is studied. The competition is introduced through the assumption that
the reproduction rate depends on the crowding in the neighborhood.
The spatial dynamics corresponds either to normal diffusion characterized by
Gaussian jumps or to superdiffusion characterized by L\'evy flights.
It is observed that in both cases periodic patterns occur for appropriate parameters
of the model, indicating that the general macroscopic collective behavior of
the system is more strongly influenced by the competition for the resources than
by the type of spatial dynamics. However, some differences arise that are discussed.
}

\begin{document}

\maketitle


\section{Introduction}

Interacting particle systems help to model and understand various problems in many diverse fields.
In biological contexts they are particularly important to study aggregation phenomena of individuals.
Fish schools, insect swarms, bacterial patterns, bird flocks and patchy plankton structures are just a
few examples revealing the ubiquity and fundamental importance of organism aggregates.

An attempt to address a simple mechanism giving rise to the
clustering of particles (with emphasis on plankton patchiness)
was made within a Brownian bug model \cite{Young2001} (see also
refs.~\cite{Zhang1990,Felsenstein1975}). The model consists of
an ensemble of particles (bugs), each one dying or reproducing
with a given probability and undergoing Brownian motion. If the
diffusivity of the particles is low enough, {macroscopic
spatial clustering occurs, since a newborn is located close to
the parent but particles can die anywhere. If diffusivity is
large, particles perform more extended walks and the region
that was left empty due to the death of a particle, is occupied
fast by some other particle. Similar results were obtained in
refs.~\cite{BHpre2002, BHprl2008, BHpre2009, PS2004} where
lattice-models were studied.}

The basic Brownian bug model lacks any interaction between the
particles. In a more realistic model (interacting Brownian bug
model) the inter-particle interaction was taken into account
assuming that the birth and death of individuals depend on the
number of other bugs in the neighborhood
\cite{Hernandez2004,Hernandez2005,Lopez2004,Bolkel1999,Martin2004,Birch2006}.
For appropriate parameters, a salient property of that model is
the formation of spatially periodic clustering of bugs
\cite{Hernandez2004}. For large diffusion the clusters become
blurred and the periodic pattern is replaced by a {more uniform
distribution of bugs. Importantly, whereas the positive
correlations leading to clustering in the non-interacting bug
case arise from the reproductive correlations, {\it i.e.}, from
the fact that offspring is born at the same location of parent
\cite{BHprl2008, BHpre2009}, the periodic arrangement of
clusters in the system of interacting bugs is a consequence of
the competitive interaction and has a spatial scale determined
by the interaction range $R$ \cite{Hernandez2004}.}

At the same time it has been observed that many living organisms move consistently with L\'evy flight behavior
\cite{metzler2000,Metzler2004,Dieterich2008,Klages2008,Viswanathan2008}.
In particular, the motion of some bacteria is found to be described by L\'evy statistics \cite{levandowsky1997,
nossal1983}, as well as the movement of spider monkeys in search of food \cite{spiders}.
The L\'evy type of motion has been shown to be advantageous with respect to standard Brownian
motion in some searching strategies involving foraging \cite{Viswanathan2008}, or in order to enhance encounter rates
at low densities \cite{Bartumeus2003}.
The main reason for this resides in the occurrence of occasional long jumps.

However, the impact of L\'evy-type diffusion on the properties
of organism aggregates has not received much attention thus
far. In the present paper we investigate in the context of a
simple interacting bug model, {continuous in space}, how the
occurrence of long jumps in the motion of the individuals
influences the collective behavior (c.f.
ref.~\cite{Durang2009}). Similarities and differences between
the interacting Brownian and L\'evy bug systems will be
highlighted.


\section{Model and numerical algorithm} \label{sec-model}


We consider a system consisting initially of $N_0$ pointlike
particles, placed randomly in a two-dimensional $L \times L$
square domain with periodic boundary conditions. {After the
random time $\tau$, a particle $i$, chosen randomly among all
the $N(t)$ bugs in the system at the present time $t$,
undergoes one of the two events: it either reproduces or
disappears. For the birth and death rates of the $i$-th
particle we assume \cite{Hernandez2004},
\begin{eqnarray}
r^i_b &=& \mathrm{max} \left( 0, r_{b0} - \alpha N_R^i \right) \, , \label{r-birth} \\
r^i_d &=& \mathrm{max} \left( 0, r_{d0} + \beta N_R^i \right) \, . \label{r-death}
\end{eqnarray}
Here $N_R^i$ is the number of particles which are at a distance
smaller than $R$ from particle $i$, the parameters $r_{b0}$ and
$r_{d0}$ are the zero-density birth and death rates, and the
parameters $\alpha$ and $\beta$ determine how $r^i_b$ and
$r^i_d$ depend on the neighborhood; the function $\mathrm{max}()$ enforces the positivity of the rates.
Such a choice for the reproduction and death rates introduces an interaction between the bugs.
For the random times $\tau$ an exponential probability density with the complementary cumulative distribution
\begin{equation} \label{Ptau-exp}
p(\tau) = \exp(-\tau / \tilde{\tau})
\end{equation}
and a characteristic time $\langle \tau \rangle = \tilde{\tau} =  R_\mathrm{tot}^{-1}$ is chosen;
$\tilde{\tau}$ is the time-scale parameter and
\begin{equation} \label{R-tot}
R_\mathrm{tot} = \sum _{i=1} ^N (r^i_b + r^i_d)
\end{equation}
is the total rate of all the $N \equiv N(t)$ particles.
In the case of reproduction, the new bug is located at the same position $(x_i, y_i)$ as the parent particle $i$.
After the demographic event, {\it i.e.}, after each random time $\tau$, all the particles perform a jump of random length $\ell$ in a random direction characterized by an angle uniformly distributed on $(0, 2 \pi)$ ($\ell$ and the direction of the jump are different for each particle).
The new present time is  $t + \tau$ and the process is repeated again, until the final simulation time is reached \cite{Birch2006}.

In order to simulate the system where the particles undergo normal diffusion, a Gaussian jump-length probability density
function is used,
\begin{equation}
\varphi(\ell) = \left (\tilde{\ell} \sqrt{2\pi} \right )^{-1}
\exp \left[ -\ell ^2 / (2 {\tilde{\ell}}^2) \right] \, ,
\end{equation}
with variance $\langle \ell ^2 \rangle = {\tilde{\ell}}^2$; $\tilde{\ell}$ is the space-scale parameter.
Since we draw the angle specifying the direction of the jump from the
interval $(0,2\pi)$, we can neglect the sign of $\ell$.
Note that the random walk defined in this way is not exactly the same
as the one in which the walker performs jumps
extracted from a two-dimensional Gaussian distribution, but it
also leads to normal diffusion and allows a more direct comparison with the L\'evy case.
The corresponding diffusion coefficient is
\begin{equation} \label{kappa}
\kappa = \langle \ell ^2 \rangle / (2 \langle \tau \rangle) \, ,
\end{equation}
which expressed through  the space- and time-scale parameters
reads, {$ \kappa = {\tilde{\ell}}^2 / (2 \tilde{\tau})$.} As we
choose to fix the value of $\kappa$, then the space-scale
parameter is determined by {$ \tilde{\ell} = \sqrt{2 \kappa
\tilde{\tau}} = \sqrt{2 \kappa / R_\mathrm{tot}}$.} Note that
because the number of particles is changing in time also the
quantity $R_\mathrm{tot}$ is changing in time. However, as in
this paper we investigate asymptotic statistically steady
states, the number of particles weakly fluctuates around a
well-defined mean value. Thus, the space-scale parameter
$\tilde{\ell}$ is fluctuating in time, but this time dependence
is not affecting qualitatively the dynamics of the system.

In order to simulate the system where the particles undergo
superdiffusion one can use a symmetric L\'evy stable probability
density function for the jump size, which behaves
asymptotically as \cite{Klages2008,metzler2000}
\begin{equation} \label{levyPDF}
\varphi _\mu (\ell) \approx {\tilde{\ell}}^{\mu}  |\ell|^{-\mu - 1} \, ,
\quad \ell \to \pm \infty \quad (|\ell| \gg \tilde{\ell}) \, ,
\end{equation}
with the L\'evy index $0 < \mu < 2$.
For all L\'evy stable probability density functions with $\mu < 2$ the variance
diverges, $\langle \ell ^2 \rangle = \infty$, leading to the
occurrence of extremely long jumps, and typical trajectories
are self-similar, on all scales showing clusters of shorter
jumps intersparsed by long excursions.
For $0 < a < \mu < 2$ fractional moments converge, $\langle |\ell|^a \rangle < \infty$.
Therefore, for the L\'evy index in the range $1 < \mu < 2$ the
value of $\langle | \ell | \rangle$ is finite, whereas for $0 < \mu \le 1$ it diverges.
The complementary cumulative distribution corresponding to (\ref{levyPDF}) is
\begin{equation} \label{Plevy}
P_\mu (\ell) \approx {\mu}^{-1} ( |\ell | / \tilde{\ell} )^{-\mu} \, ,
\quad \ell \to \pm \infty \, .
\end{equation}
As a simple form of complementary cumulative distribution function, which is normalizable and behaves asymptotically as (\ref{Plevy}), we use
\begin{equation} \label{Plambda-pareto}
P_\mu (\ell) = (1 + b^{1/\mu} |\ell | / \tilde{\ell})^{-\mu} \, ,
\end{equation}
where $b = [\Gamma(1-\mu / 2) \Gamma (\mu / 2)] / \Gamma (\mu)$.
As before, we can neglect the sign of $\ell$ since the direction of the jump is assigned by
drawing an angle on $(0,2\pi)$.
Now the diffusion coefficient (\ref{kappa}) is infinite, but
one can define a generalized diffusion coefficient in terms of
the scales $\tilde{\ell}$ and $\tilde{\tau}$ as
\cite{metzler2000,Klages2008}
\begin{equation}
\kappa_\mu = {\tilde{\ell}}^\mu / (2 \tilde{\tau}) \, .
\end{equation}
Therefore, in the case of the L\'evy flights, when fixing the value of $\kappa_\mu$, the space-scale
parameter is, $\tilde{\ell} = (2 \kappa_\mu \tilde{\tau})^{1/\mu} = (2 \kappa_\mu / R_\mathrm{tot})^{1/\mu}$.}

The model described can be interpreted in the following way:
during a time interval $\tau$ each individual moves on average
a distance $\ell$ in some direction but only one of them reproduces or dies.
For positive values of $\alpha$ and $\beta$, the more neighbors a particle has within the radius
$R$, the smaller is the probability of reproduction and the
larger is the probability that the bug does not survive, {\it e.g.}, due to competition for resources.
The periodicity of the simulation domain represents the fact that particles can
leave and enter the observation area. In the case of long jumps
it represents the situation where there is one particle that
leaves the domain and another one that arrives from somewhere,
perhaps from far away, and its position is basically random in the domain.
In principle, one could also simulate the system so that after the inter-event time $\tau$ a randomly chosen
particle undergoes one of the three events: reproduction, death, or jump.
However, the results obtained are the same as following the procedure described above; one would just have
one more parameter, the jump rate, and different numerical
values for the other parameters.

\begin{figure}[t]
\centering
\includegraphics[width=7.0cm]{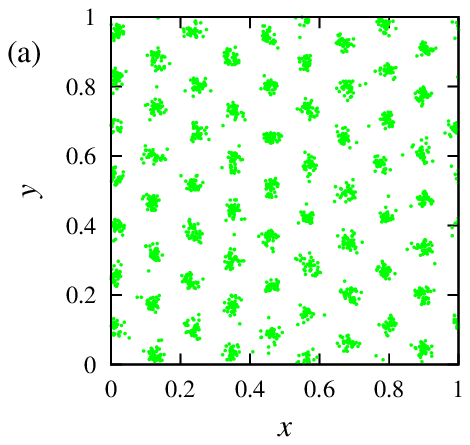} \\
\includegraphics[width=7.0cm]{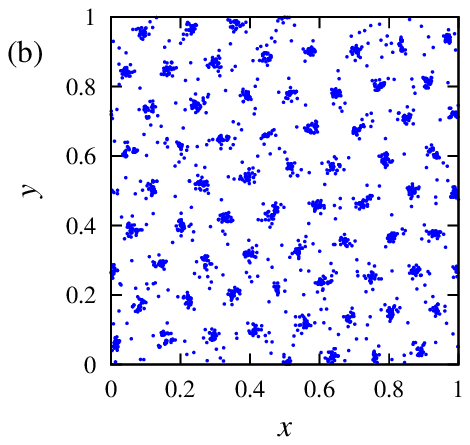}
\caption{
Interacting Brownian bugs (a) {\it versus} interacting L\'evy bugs (b): spatial
configuration of particles in the statistically steady state.
The parameters are: $r_{b0} = 1$, $r_{d0} = 0.1$, $\alpha = 0.02$, $\beta = 0$, and $R = 0.1$
(see also text).
In the case of Brownian bugs the diffusion coefficient is $\kappa = 10^{-5}$.
In the case of L\'evy bugs the value of L\'evy index is $\mu =1$ and the anomalous
diffusion coefficient is $\kappa_\mu = 56 \times 10^{-5}$.
The average number of particles in the two systems is approximately the same,
$\langle N \rangle = 2555$ and
$\langle N \rangle = 2565$, respectively.}
\label{FigILB-XY}
\end{figure}
%


\section{Results}


In the following we set $\beta = 0$ in eq.~(\ref{r-death}) and $\alpha = 0.02$ in eq.~(\ref{r-birth}), {\it i.e.}, the
probability of death is constant and the same for all the particles while the probability for the reproduction depends on
how crowded the environment is.
Also, we use the values $r_{b0} = 1$, $r_{d0} = 0.1$, $L = 1$, and $R=0.1$ in
all the figures presented in the current paper.
For $\beta=0$ the critical number of neighbors, $N_R^*$, for which death and reproduction are equally probable for particle $i$, is determined by
\begin{equation} \label{Neq}
N_R^* = \Delta _{bd}/\alpha \, ,
\end{equation}
where $\Delta _{bd} = r_{b0} - r_{d0}$.
If $N_R^i < N_R^*$ it is more probable that particle $i$
reproduces and if $N_R^i > N_R^*$ death is more likely.
In the following we fix the value of the L\'evy index to $\mu = 1$.
The detailed discussion of the influence of the birth and death
rates as well as the L\'evy index on the results is outside the
scope of the current paper and will be presented elsewhere.

\begin{figure}[t]
\centering
\includegraphics[width=7.0cm]{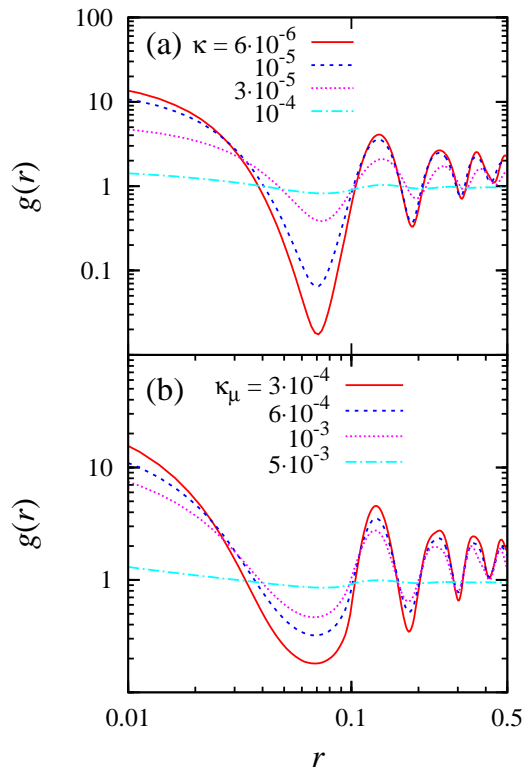}
\caption{
Radial distribution function for the system with (a) Gaussian jumps and
(b) L\'evy jumps for several values of the diffusion and generalized diffusion
coefficient, respectively.
The other parameters are as in fig.~\ref{FigILB-XY} (see also text).
}
\label{FigRad}
\end{figure}

In the case of interacting Brownian bugs, for small enough
$\kappa$ and large enough $\Delta _{bd}$, the occurrence of
periodic patterns has been observed previously
\cite{Hernandez2004, Lopez2004}. In the statistically steady
state clusters form and arrange in a hexagonal lattice (see
fig.~\ref{FigILB-XY}-a). For large values of the diffusion
coefficient {the periodic pattern is replaced by an almost
homogeneous distribution of particles} (see also
fig.~\ref{FigRad}-a). In the case of L\'evy flights, since the
diffusion coefficient (\ref{kappa}) is infinite, one could
expect that for the interacting L\'evy bugs the spatial
distribution {will not reveal a periodic pattern}. However, as
can be seen from fig.~\ref{FigILB-XY}-b, this is not the case.
The reason for the divergence of the diffusion coefficient in
the L\'{e}vy case is in the statistical weight of the large
jumps. These large jumps have some influence in the
characteristics of the patterns formed, but the large-scale
structure is ruled mainly by the interactions between
particles.

For the interacting bug system with Gaussian jumps and moderate
diffusion the appearance of periodic clustering is well
captured by combining the effects of diffusion and interaction
in a mean-field approach \cite{Hernandez2004}.
Following the steps in ref.~\cite{Hernandez2004} one obtains the following equation as a mean-field approximation to
the dynamics of the density of particles $\rho(\textbf{x},t)$
in the interacting L\'evy bug model ($\beta=0$):
\begin{equation}
\frac{\partial\rho(\textbf{x},t)}{\partial t}
= \rho(\textbf{x}, t) \left( \Delta_{bd} - \alpha  \int_{D} d \textbf{y} \rho(\textbf{y}, t) \right)
+ \kappa_\mu \nabla^\gamma \rho(\textbf{x}, t) \, \ .
\label{mean-field}
\end{equation}
The integration domain $D$ is the set of points within a
distance smaller than $R$ from $\textbf{x}$:  $\vert
\textbf{x}-\textbf{y} \vert < R$. The operator $\nabla^\gamma$,
with $\gamma = \mu$\footnote{This expression was incorrect in
the paper originally published in EPL 92, 40011 (2010); it was
corrected to the present form in the Erratum in EPL 95, 69902
(2011).}, is the fractional diffusion operator associated to
the L\'{e}vy flights of exponent $\mu \in (0, 2)$
\cite{metzler2000, Klages2008}. It reduces to the standard
diffusion operator for $\mu > 2$. {In this mean-field
description the first term accounts for the net growth of the
population, the second one takes into account the non-local
contribution associated to the saturation due to the
interactions within a distance $R$, and the third term
describes the spatial diffusion of particles.
Equation~(\ref{mean-field}) differs from the mean-field
approximation derived in ref.~\cite{Hernandez2004} only in the
third term where the standard diffusion operator is replaced by
the fractional diffusion operator and the diffusion coefficient
by the generalized diffusion coefficient. Similar descriptions
of reaction-diffusion systems with a fractional diffusion term
can be consulted, {\it e.g.}, in refs.~\cite{Baeumer2007,
Negrete2003, metzler2000, Metzler2004}.}

\begin{figure}[t]
\centering
\includegraphics[width=7.0cm]{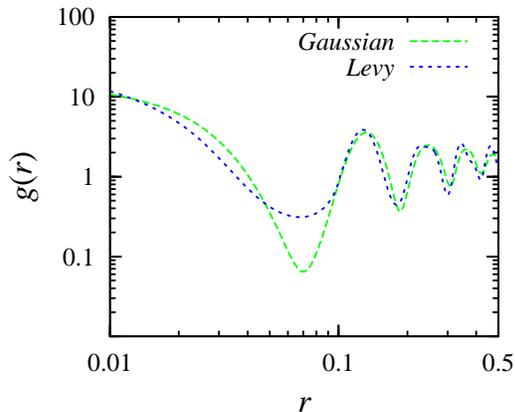}
\caption{
Comparison of the radial distribution functions for the systems with Gaussian
jumps and with L\'evy flights.
The parameters are as in fig.~\ref{FigILB-XY}.
}
\label{FigRad-GP}
\end{figure}

Equation~(\ref{mean-field}) has always the uniform solution
\begin{equation}
\rho (\textbf{x}, t) = \rho_0 = \Delta_{bd} / (\alpha \pi R^2) \, ,
\end{equation}
which turns out to become unstable for small $\kappa_\mu$ and/or
large $\Delta_{bd}$ (what is small or large depends also on the
value of $\mu$).
This can be seen by introducing the {\sl ansatz} $\rho (\textbf{x}, t) = \rho_0 + \delta \rho(\textbf{x}, t)$ in eq.~(\ref{mean-field}) and linearizing
in $\delta \rho (\textbf{x}, t)$.
The result is
\begin{equation}
\delta \rho (\textbf{x}, t) \sim \exp (i \textbf{k}\cdot\textbf{x} + \lambda(\textbf{k}) t) \, ,
\end{equation}
with
\begin{equation} \label{LWN}
\lambda(|\textbf{k}|) = - \kappa_\mu | \textbf{k} | ^\gamma - 2 \Delta_{bd} J_1 (|\textbf{k}|R) / (|\textbf{k}|R) \, ;
\end{equation}
$J_1$ is a Bessel function. {An instability of the uniform
solution will occur if the sign of $\lambda(|{\bf k}|)$ changes
from negative to positive at some value of $|{\bf k}|$. This
will occur first for the critical wavenumber $|{\bf k}_c|$ for
which $\lambda(|{\bf k}|)$ is maximum. The instability will
develop in a periodic pattern which, at least close enough to
the instability, will have a periodicity $\delta = 2\pi/|{\bf
k}_c|$. Introducing a dimensionless wavenumber $q \equiv |{\bf
k}|R$ and growth rate $\Lambda = R^\gamma\lambda/\kappa_\mu$,
eq.~(\ref{LWN}) reads,
\begin{equation}
\Lambda(q)= -q^\gamma-\nu J_1(q) / q \ .
\end{equation}
The latter equation shows that, within this mean-field description, the relevant parameters are $\gamma$ (or $\mu$) and the
dimensionless quantity $\nu \equiv 2R^\gamma\Delta_{ab}/\kappa_\mu$.
Imposing the condition $\Lambda(q)=0$, corresponding to the change of sign in the growth rate,
and $\Lambda'(q) = 0$, corresponding to the maximum of the growth rate,
one can find the critical value of $\nu$, $\nu_c=-q_c^{\gamma+1}/J_1(q_c)$,
and the equation for the critical wavenumber:
\begin{equation}
q_c J_2(q_c) / J_1(q_c) = - \gamma \, .
\end{equation}
Numerical solution of the latter equation provides $q_c$, and therefore $\delta$, as a function of $\gamma$ (or $\mu$), and that $\delta$ is proportional to $R$.
In order to make a comparison with the results from numerical simulations of the system, one should keep in mind that the periodic boundary
conditions will change $q_c$ to the closest number of the form $(2\pi R/L) \sqrt{n^2+m^2}$ with $n$ and $m$ integers. For $\mu =1$ the value of $\delta$ obtained from the mean-field approximation is $0.128036$ and for $\mu = 2$ (Gaussian case) it is $0.131306$, {\it i.e.}, the periodicity of the pattern is in both cases of the order of $R = 0.1$.
}

{To extract the periodicity of the pattern from the numerical
simulations of the system, it is useful to study the radial
distribution function $g(r)$.} It describes how the density
varies with the distance from a given particle respect to the
one expected from a uniform distribution, giving thus
additional information about the distribution of bugs
\cite{Ramos2008}. It is computed in the standard way, {\it
i.e.}, by counting all particles, $dn$, at a distance between
$r$ and $r+dr$ from the target particle, using the formula $dn
= (N/L^2) g(r) 2 \pi r dr$, and averaging over all particles
and different long times. In fig.~\ref{FigRad-GP} the
comparison of the radial distribution functions for the
Brownian and L\'evy bugs (for the same systems as in
fig.~\ref{FigILB-XY}) is depicted. Figure~\ref{FigRad}-a shows
the behavior of $g(r)$ in the case of Brownian bugs and in
fig.~\ref{FigRad}-b in the case of L\'evy bugs for various
values of diffusion coefficient and generalized diffusion
coefficient, respectively. {The second maximum of $g(r)$,
indicating the periodicity of the pattern, appears in
fig.~\ref{FigRad}-a (in the systems of Brownian bugs), up to
$\kappa = 10^{-5}$, at $0.13125$ and in fig.~\ref{FigRad}-b (in
the L\'evy case), up to $\kappa_\mu = 10^{-3}$,  at $0.12875$,
being in good agreement with the results obtained from the
mean-fieled approximation, and becoming slightly larger only at
larger values of $\kappa$ and $\kappa_\mu$. The anomalous
exponent $\mu$ has only a very light influence on the
periodicity of the pattern, which is of the order of $R$ (see
also fig.~\ref{FigRad-GP} where the slight shift can be can be
noticed). }

Figure~\ref{FigRad} shows also that as $\kappa$ or
$\kappa_{\mu}$ increases, the first peak of $g(r)$ gets lower
and wider: the total number of particles in the system
decreases and the clusters become more spread. For large values
of diffusivity ($\kappa$ or $\kappa _\mu$), {the oscillations
of $g(r)$ smooth out, indicating that the periodic pattern
becomes replaced by a more homogeneous distribution of bugs.}

\begin{figure}[t]
\centering
\includegraphics[width=7.0cm]{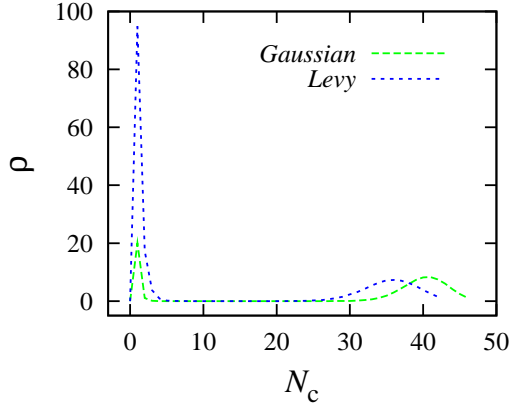}
\caption{
The distribution of cluster sizes for the interacting Brownian bug and L\'evy bug models.
All parameters have the same values as in figs.~\ref{FigILB-XY} and \ref{FigRad-GP}.
}
\label{Fig-clustersize}
\end{figure}
\begin{figure}[t]
\centering
\includegraphics[width=7.0cm]{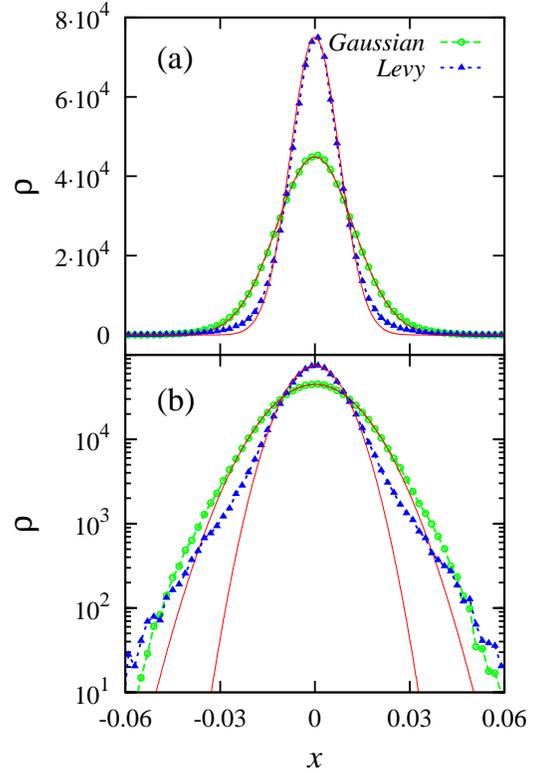}
\caption{
Cross-section of the two-dimensional particle density of the average cluster in
(a) linear and (b) semi-log scale.
Comparison between the interacting Brownian and L\'evy bug models.
The values of the parameters are the same as in figs.~\ref{FigILB-XY}, \ref{FigRad-GP}, and \ref{Fig-clustersize}.
The continuous lines correspond to the fitting with Gaussian functions.}
\label{Fig-clustershape}
\end{figure}

As a difference compared to the case of interacting Brownian bugs, we observe that now, even at small values of $\kappa_\mu$,
there are many solitary particles appearing for short time periods in the space between the periodically arranged clusters,
due to the large jumps, c.f. figs.~\ref{FigILB-XY}-a and \ref{FigILB-XY}-b.
This is even better illustrated by fig.~\ref{Fig-clustersize} where the probability distributions
of the cluster sizes for the systems with Gaussian and L\'evy
jumps are depicted (for the same systems as in
fig.~\ref{FigILB-XY}, {\it i.e.}, for the given parameters the
values of $\kappa$ and $\kappa _ \mu$ have been chosen such
that the average number of the particles in the two systems is
approximately equal).
The clusters are defined using the nearest neighbor clustering method \cite{florek1951}, with a
threshold value 0.025 to define neighbors.
As one can see, for the system of interacting Brownian bugs the most probable size of large clusters
forming the periodic pattern is $41$ particles whereas for
interacting L\'evy bugs it is $36$ (the critical number of neighbors is in both cases $N_R^* = 45$).
A noticeable difference occurs in the proportion of single-particle clusters: the peak of the
distribution at the value characterizing isolated particles
($N_c = 1$) is much higher in the L\'evy flight case.
Returning to fig.~\ref{FigRad-GP}, which refers to the same systems as in figs.~\ref{FigILB-XY} and \ref{Fig-clustersize}, it is to be noticed that the minimum between
the first and the second peak is much lower for the system of Brownian bugs.
This result is consistent with fig.~\ref{Fig-clustersize}: in the system with L\'evy flights, the occurrence of many single particles between the
clusters forming the periodic pattern produces a higher inter-cluster density is observed.

Finally, in fig.~\ref{Fig-clustershape} the cross-section of the two-dimensional
particle density of the average cluster is depicted for the interacting Brownian and L\'evy bug systems (same systems as in fig.~\ref{FigILB-XY}).
The average cluster is obtained by setting the origin at the center
of mass of each cluster forming the periodic pattern and
averaging over all the clusters in the simulation area and over time.
In the case of Brownian as well as of L\'evy bugs, the central part of the average cluster, where most of the
particles are concentrated, is well fitted by a Gaussian function (continuous curve).
However, fig.~\ref{Fig-clustershape}-b reveals the difference in the way
the particle density decreases in the two systems when moving
away from the center of mass of the cluster. Though even in the
case of the Brownian bugs the tail of the average cluster is
not really well described by a Gaussian function, a Gaussian
decay provides a good approximation. Instead, in the case of
the L\'evy bugs the tail of the average cluster decays slower.
Since the average cluster is calculated from individual clusters which are separated by a distance of order $R$,
one cannot properly estimate an asymptotic decay, but the decay in the L\'{e}vy case seems close to exponential.
In any case, it is distinctively faster than the power-law behavior which would be exhibited by
clusters of non-interacting particles moving purely by L\'{e}vy flights.
The existence of the interaction range $R$ introduces a cut-off distance which makes the long-jumps characteristic of
L\'{e}vy flights not so relevant to determine the large-scale properties of the spatial particle configurations.


\section{Conclusion}
\label{conclusion}

In the present paper we have studied a simple birth-death model
with competition among the individuals of the same species. In
particular, we have investigated how the superdiffusive motion
of the individuals characterized by L\'evy flights influences
the collective behavior. We have observed that the appearance
of {periodic} clustering known from previous studies of the
interacting Brownian bug model takes place also in the
interacting L\'evy bug model for appropriate parameters. This
is against the expectation that particles performing L\'evy
flights cannot give rise to {space-periodic clustering} since
their diffusion coefficient is infinite, for which the
interacting Brownian bug model is known to reveal {no periodic
pattern}. However, as a difference we have observed that, in
the interacting L\'evy bug model, due to the long jumps, there
are many single particles between the clusters, leading also to
the differences in the particle density profiles of the average
cluster or in the short-distance behavior of the radial
distribution function. From this one can conclude that the
large-scale collective behavior of the system is much more
strongly influenced by the competitive interaction than by the
type of spatial motion performed by the bugs. {We have also
verified that the mean-field approximation (\ref{mean-field})
is proper to describe the periodicity in the interacting L\'evy
bug model.

As a final remark, let us mention that though the model studied
in the present article describes rather living organisms, such
as animals or bacteria, non-local interactions and L\'evy
flights are important also in plant ecology for the development
of vegetation patterns \cite{Borgogno2009}, due to the
competition for the resources and because the seed dispersal is
often described better by L\'evy flights than by Gaussian jumps
\cite{Clark1999}.}


\acknowledgments
This work has been supported by the targeted
financing project SF0690030s09, Estonian Science Foundation
through grant no. 7466, by the Balearic Government (E.H.),
and by Spanish MICINN and FEDER through project FISICOS
(FIS2007-60327).



\bibliography{Refs-Heinsalu}
\bibliographystyle{eplbib}

\end{document}